%% file: main.tex
\documentclass[sigconf]{acmart}

\usepackage{multirow}
\usepackage{algpseudocode}
\usepackage{algorithm}
\AtBeginDocument{%
  }

\copyrightyear{2022}
\acmYear{2022}
\setcopyright{acmcopyright}\acmConference[ASE '22]{37th IEEE/ACM International Conference on Automated Software Engineering}{October 10--14, 2022}{Rochester, MI, USA}
\acmBooktitle{37th IEEE/ACM International Conference on Automated Software Engineering (ASE '22), October 10--14, 2022, Rochester, MI, USA}
\acmPrice{15.00}
\acmDOI{10.1145/3551349.3556909}
\acmISBN{978-1-4503-9475-8/22/10}

\begin{document}

\title{Low-Resources Project-Specific Code Summarization}


\thanks{$\dagger$ Both authors contributed equally to this research.}
\thanks{$*$ Corresponding author.}

\author{Rui Xie$^{\dagger}$}
\author{Tianxiang Hu$^{\dagger}$}
\author{Wei Ye$^{*}$}
\author{Shikun Zhang$^{*}$}
\email{ruixie, hutianxiang, wye, shangsk@pku.edu.cn}
\affiliation{
  \institution{National Engineering Research Center for Software Engineering, Peking University}
  \city{Beijing}
  \country{China}
}

\begin{abstract}
Code summarization generates brief natural language descriptions of source code pieces, which can assist developers in understanding code and reduce documentation workload. 
Recent neural models on code summarization are trained and evaluated on large-scale multi-project datasets consisting of independent code-summary pairs. 
Despite the technical advances, their effectiveness on a specific project is rarely explored. 
In practical scenarios, however, developers are more concerned with generating high-quality summaries for their working projects. 
And these projects may not maintain sufficient documentation, hence having few historical code-summary pairs.
To this end, we investigate low-resource project-specific code summarization, a novel task more consistent with the developers' requirements. 
To better characterize project-specific knowledge with limited training samples, we propose a meta transfer learning method by incorporating a lightweight fine-tuning mechanism into a meta-learning framework. 
Experimental results on nine real-world projects verify the superiority of our method over alternative ones and reveal how the project-specific knowledge is learned.
\end{abstract}


\maketitle

\input{intro}
\input{approach}
\input{experiment}

\input{results}

\input{discussion}

\input{related_work}
\input{conclusion}

\begin{acks}
This research was supported by the China Postdoctoral Science Foundation(No. 2021M700216). Any opinions, findings, and conclusions expressed herein are the authors and do not necessarily reflect those of the sponsors.
\end{acks}

\bibliographystyle{ACM-Reference-Format}
\bibliography{bib}

\end{document}

%% file: intro.tex
\section{Introduction} \label{sec:intro}

Code summaries, also referred to as code documentation, are readable natural language texts that describe source code's functionality and serve as one of the most common support to help developers understand programs~\cite{haiduc2010supporting}. Data-driven automatic code summarization has now become a rapidly-growing research topic.
Researchers in the software engineering and natural language processing communities have proposed a variety of neural models for code summarization. These models are usually built upon techniques widely used in machine translation and text summarization, such as reinforcement learning~\cite{Yao2018Improving}, Variational AutoEncoders~\cite{Chen2018VAE}, dual learning~\cite{DualCos}\cite{ye2020dual}, and retrieval techniques~\cite{gregg2020rencos, li2021EditSum}.

Previous neural code summarization models are typically trained and evaluated on large-scale datasets consisting of independent code-summary pairs from many software projects. They are referred as \textbf{\textit{general code summarization (GCS)}} models in this paper. Despite the promising results of recent GCS methods, few of them explore their effectiveness on a specific project, which, however, is more concerned by developers in practical scenarios. After all, what developers need more is not a performant model over cross-project datasets, but a tool to generate high-quality and consistent code summaries for their specific working projects. We term the scenario of generating code summaries for an individual project \textbf{\textit{project-specific code summarization (PCS)}}.



\begin{table}[t]
  \caption{
  Motivation Scenario. Current code summarization models tend to generate summaries in the manner used by most projects  (e.g., with the frequent pattern of ``return true if'')  rather than the way of the target project (e.g., with the pattern of ``checks whether'' of project Flink). Meanwhile, directly applying a code summarization model in specific projects may generate semantically-poor summaries due to the lack of project-specific domain knowledge. The summaries in this example are generated by a Transformer-based code summarization model~\cite{wasi2020transformer}  trained on a large-scale dataset proposed by LeClair\cite{Leclair2019recommendations}.
  }
  \label{tab:summary_patterns}
  \centering
  \begin{tabular}{cl}
    \hline
        \multicolumn{2}{c}{\textbf{Guava}} \\
    \hline
        Source Code & public boolean contains(...) \{ \\
        & \, final Monitor monitor = this.monitor; \\
        & \, monitor.enter(); \\
        & \, ...\\
        & \, return q.contains(o); \\
        & \} \\
    \hline
        Human-Written & \textbf{returns true if} this queue contains the ... \\
    \hline
        Transformer & \textbf{returns tt true tt if} this multimap contains ... \\
    \hline
        \multicolumn{2}{c}{\textbf{Flink}} \\
    \hline
        Source Code & public boolean isEmpty() \{ \\
        & \, return size() == 0; \\
        & \} \\
    \hline
        Human-Written & \textbf{checks whether} the queue is empty has no ... \\
    \hline
        Transformer & \textbf{returns true} if this map contains no key ... \\
    \hline
\end{tabular}
\end{table}

Unfortunately, a good GCS model is not guaranteed to be a good PCS one. The reason mainly lies in that GCS models typically focus on capturing common cross-project semantics, potentially leading the generated summaries to fail to capture project-specific characteristics.  For example, one software project has its unique domain knowledge, and in most cases, its documentation is written in a relatively consistent style.  Under the task settings of GCS, however, models tend to generate monotonous summaries in the manner used by most projects,  rather than the way or style of the target project. 
Table \ref{tab:summary_patterns} shows a concrete case, where a robust Transformer-based GCS model \cite{wasi2020transformer}  is trained on a large-scale multi-project dataset \cite{Leclair2019recommendations}. For the two code snippets in Table \ref{tab:summary_patterns}, Transformer generates code summaries with an expression style of "return true if." These summaries are consistent with the finding of Li et al. \cite{li2021EditSum} that the pattern of ``returns true if'' frequently appears in the dataset provided by LeClair et al. \cite{Leclair2019recommendations}. The problem here is that project Flink prefers another writing style of ``check whether'', making the generated summary incoherent with its historical summaries. Meanwhile, we can easily find that the generated summaries do not cover enough meaningful topic words, since a GCS model may have limited domain knowledge of specific projects (Guava and Flink here).
These facts inspire us that project-specific knowledge should be better distilled to improve code summary quality in PCS.



A natural technical design for PCS models is to introduce transfer learning, by considering each project as a unique domain. In our preliminary experiment,  a classical fine-tuning strategy yields robust performance improvements in projects with numerous off-the-shelf code summaries. However, going back to the practical development scenario again, we find that projects may usually be poorly documented,  with insufficient code summaries for fine-tuning. According to our investigation of 300 projects from three prominent open-source organizations (Apache, Google, and Spring),  a large number of the projects lack historical code summaries in terms of model training. For example,  as shown in Table \ref{tab:stat_projects}, nearly one-third of projects have code summaries of less than 100. 
Considering the authority of these three organizations, we believe that there are more open-source projects with a small number or even none of existing code summaries. 
To this end, this paper proposes a novel and essential task—low-resource project-specific code summarization—to tackle a more practical code summarization challenge in the software engineering community. 

\begin{table}[t]
  \caption{Statistics of the number of open-source projects whose code summary counts are less than 10 or 100. 
  We selected 100 projects from open-source organizations of Apache, Google, and Spring, respectively. 
  And then we statistic the number of summaries for public methods.}
  \label{tab:stat_projects}
  \centering
  \begin{tabular}{lccc }
  \hline
    \multirow{2}*{Source}& \multirow{2}*{Total} & \multicolumn{2}{c}{ Number of Projects } \\
    & & \#Summary \textless 10 & \#Summary \textless 100 \\
    \hline
      Apache  & 100 & 3 & 11 \\ 
      Google  & 100 & 11 & 41 \\
      Spring  & 100 & 13 & 44 \\
    \hline  
      Total   & 300 & 27 & 96 \\
    \hline
\end{tabular}
\end{table}

As a pioneering effort on low-resource project-specific code summarization, we then propose a simple yet effective meta-learning-based approach with the following two characteristics.

1. Since one development organization usually has more than one working project, we investigate how to leverage multiple projects to promote transferring project-specific knowledge. Unlike conventional GCS methods that uniformly treat data samples from different projects, our method regards each project as an independent domain to better characterize project features. Specifically, due to the efficacy of meta-learning in handling low-resource applications\cite{ChenS21a}, we introduce Model-Agnostic Meta-Learning (MAML) \cite{chelsea2017maml} into PCS. Meta transfer learning in our scenario means learning to do transfer learning via fine-tuning multiple projects together. More specifically, we condense shared cross-project knowledge into the form of weight initialization of neural models, enabling better project-specific domain adaption.

2. Code summarization models utilizing modern sequence-to-sequence architecture usually have large-scale parameters, bringing two limitations to PCS. On the one hand, project-specific knowledge is accumulated and updated frequently with code or documentation revisions, hindering the efficiency of iterative model optimization. On the other hand, limited training data for a specific project will easily make big models overfitted. Therefore, inspired by the recent advent of prompt learning in the NLP community\cite{xiang2021prefix}, in our meta transfer learning, we keep the pre-trained model parameters frozen and only optimize a sequence of continuous project-specific vectors. These vectors, named project-specific prefix in our paper, only involve a small number of extra parameters, effectively improving the overall meta transfer learning process.

We curate a PCS dataset consisting of nine diversified real-world projects. 
The automatic and human evaluation results on this dataset verify the overall effectiveness of our method and the necessity of individual components, and, more importantly, suggest promising research opportunities on project-specific code summarization. 

The contributions of this paper are listed as follows:

\begin{itemize}

\item We propose low-resource project-specific code summarization, an essential and novel task more consistent with practical developing scenarios.

\item As a pioneering exploration on low-resource project-specific code summarization, we design \textbf{M}eta \textbf{P}refix-tuning for \textbf{CO}de \textbf{S}ummarization (\textbf{MPCos}). MPCos captures project knowledge effectively and efficiently, by integrating a project-specific prefix-based fine-tuning mechanism into a meta-learning framework, serving as a solid baseline for future study.

\item By looking into the token frequency patterns in our generated summaries, we reveal the internal process of project-specific knowledge learning to some extent.

\end{itemize}

%% file: approach.tex
\section{Problem Formulation}

We define the low-resource project-specific code summarization problem as follows. Given a target project with limited code-summary pairs corpus: $C^{\rm tgt} = \{(X_i,\hat{Y}_i)\}_{i=1..N^{\rm tgt}}$, where $N^{\rm tgt}$ is the number of code-summary pairs in the target project, the code summarization model is supposed learning how to generate correct summaries for the target project, with the help of the general knowledge of large-scale multi-project corpus and the cross-domain knowledge of few accompanying projects. Note that a target project can also serve as an accompanying one for other projects. Our task setting in this paper involves nine target projects, which means each one has eight accompanying projects. The task should care about the overall performance of all target projects instead of a single one.
In the low-resource situation, the number of code-summary pairs in target project $N^{\rm tgt}$ is usually small. Therefore, following previous work and real-world development experience, we mainly focus on two settings where $N^{\rm tgt}=10$ and $N^{\rm tgt}=100$.
\section{Methodologies}

\begin{figure*}[t]
\includegraphics[width=1.0\linewidth]{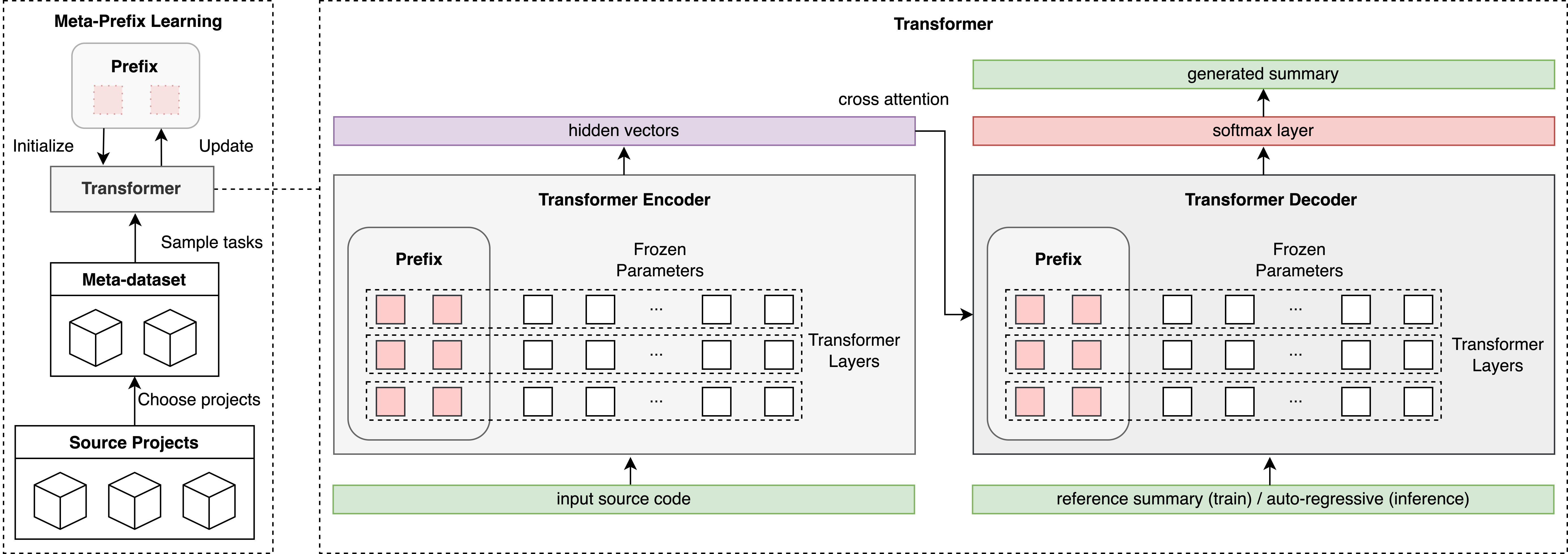}
\caption{The overall framework of our method. We first obtain the initial weights of the project-specific prefix vectors via meta training on meta datasets. Then we use these weights to initialize the model of the target project (e.g., Flink) to be fine-tuned. Finally, we train the target model on the target dataset via prefix-tuning.}
\label{fig:model}
\Description{}
\end{figure*}


We employ a classical Transformer as the backbone of our method. As shown in Figure \ref{fig:model}, MPCos mainly consists of three components: (1) Code Summarization Module, which generates target summaries based on the Encoder-Decoder architecture of Transformer; (2) Meta-Transfer Learning Module, which leverages multiple projects to learn better initial weights for prefix-tuning; and (3) Prefix-Tuning Module, which preserves separate prefixes for each project to promote project-specific transfer learning and avoid data cross-contamination. We will describe the details of each component in the following subsection.

\subsection{Code Summarization}

In recent years, the Transformer framework has been widely used in generative tasks and achieved good results. Therefore, we use a Transformer's encoder-decoder framework as our base model. In order to process programming languages and natural languages, we first tokenize input code $X$ into a code token sequence $[x_1; ...; x_{N_X}]$ where $x_i$ represents a code token in original code, and $N_X$ represents the number of tokens; for the target summary $\hat{Y}$, we also tokenize it into a sequence of tokens $[\hat{y}_1, ..., \hat{y}_{N_{\hat{Y}}}]$ where $Y_i$ represents a summary token in target summary, and $N_{\hat{Y}}$ represents the number of tokens.

The base model consists of a Transformer encoder and a Transformer decoder. Transformer encoder takes code token sequences as input, and generate a hidden vector of the input code.

\begin{equation}
    H = {\rm TransformerEncoder}(X),
\end{equation}

Transformer encoder consists of stacked transformer layers. Each transformer layer takes last layer's output as input and use multi-head attention mechanism to enhance the represents:

\begin{equation}
\begin{aligned}
    MultiHead(q,v) &= Concat(head_1, ..., head_n) \\
    where\ head_i &= Attention(q,v)
\end{aligned}
\end{equation}

\begin{equation}
\begin{aligned}
    \hat H^{l} &= LN(MultiHead(H^{l-1}, H^{l-1}) + H^{l-1})
\end{aligned}
\end{equation}

\begin{equation}
\begin{aligned}
    H^l &= LN(FFN(\hat H^{l}) + \hat H^{l})
\end{aligned}
\end{equation}

where $l$ represents the $l_{th}$ layer in transformer encoder, n represents the number of heads, $Attention$ represents classical attention mechanism proposed by \cite{vaswani2017attention}, $LN$ represents Layer normalization, $FFN$ represents a feed-forward network, and we use the input code embeddings $emb(X)$ as the initial state of $H^{0}$.

Transformer decoder takes $H$ as input and generate summary in an auto-regressive way. At time $t$ , given hidden vector $H$ and previous generated summary $Y_{<t}=\{y_1,y_2,...,y_{t-1}\}$, the decoder will generate a hidden vector represents the generated word:

\begin{equation}
    S_t = {\rm TransformerDecoder}(Y_{<t}, H).
\end{equation}

Transformer decoder also consists of several stacked transformer layers. And the output of last layer can be further used to estimate the probability distribution of word $y_t$:

\begin{equation}
    p(y_t | X) \doteq {\rm softmax}(W^Y S_t).
\end{equation}

where $W^Y$ is a weight matrix.







We denoted the base model as $M_\theta$, where $\theta$ indicates the trainable parameters in the model. Thus, the model can used to estimate distributions of target summary as:

\begin{equation}
    p(y_t | X, \theta) \doteq M(Y_{<t}, X, \theta)
\end{equation}

To utilize general code summarization knowledge, we first train $M_\theta$ with general corpus $C^{\rm pre}$. Here, we use general code summarization dataset proposed by ~\cite{Leclair2019recommendations}.

Given an input code $X$ and its ground-truth summary $\hat{Y}=\{\hat{y}_1,\hat{y}_2,...,\hat{y}_{|\hat{Y}|}\}$ from $C^{\rm pre}$, we optimize the model to minimize the negative log-likelihood (NLL) as:

\begin{equation}
    \mathcal{L}_{\rm NLL} = -\log p(\hat{Y}; X, \theta) = -\frac{1}{|\hat{Y}|}\sum^{|\hat{Y}|}_{t=1} Pr(y_t=\hat{y}_t|X, \theta)
\end{equation}

\subsection{Prefix Tuning}


To prevent over-fitting  when training large pre-trained model on low-resource scenario, we propose restricting the number of meta-trainable parameters and layers. In particular, we apply prefix tuning to reduce trainable parameters. 

The prefix-tuning is a prompting mechanism prepending to Transformer model by inserting a prefix vector into each layer of the transformer. Taking encoder as example, the prefix-tuning Transformer layer can be expressed as:

\begin{equation}
\begin{aligned}
    P^l &= MLP^l(emb^l(project))
\end{aligned}
\end{equation}

\begin{equation}
\begin{aligned}
    \hat H^{l} &= LN(MultiHead(H^{l-1}, [P^l; H^{l-1}]) + H^{l-1})
\end{aligned}
\end{equation}

\begin{equation}
\begin{aligned}
    H^l &= LN(FFN(\hat H^{l}) + \hat H^{l})
\end{aligned}
\end{equation}
where $P^l$ represents the prefixed vectors of corresponding project on the $l_th$ layer, $emb^l$ represents the projection operation for layer $l$ from source project to its corresponding embedding based on the embedding matrix $M^l$, and $MLP^l$ represents the classical Multilayer Perceptron network for layer $l$.
Following ~\cite{xiang2021prefix}, we update the parameters of $MLP^l$ and the embedding matrix $M^l$ during training. Once training is complete, these parameters can be dropped, and only the prefixed vectors $P^l$ needs to be saved.
The illustration of the proposed prefix tuning model is shown in Figure \ref{fig:model}.


\subsection{Meta Transfer Learning}

The next step is to perform meta transfer learning for fast adaption on the target project. We insert the prefix module into the framework after pre-training the base model on $C^{\rm pre}$. The prefix parameter are initialize randomly and base model parameters $\theta$ are frozen during this process.

We first construct a collection of projects $\{\mathcal{P}_i\}$ as meta-dataset $C^{\rm meta}$. For a specific target project, we selected different projects according to certain standards, which will be described in following section.

For a meta project $\mathcal{P}_i$, it contains samples from the corresponding project, and we divide them into a support set $c^{\rm meta-sup}_i=\{(X^{\rm sup}_j,\hat{Y}^{\rm sup}_j)\}_{j=1..K}$ and a query set $c^{\rm meta-qry}_i=\{(X^{\rm qry}_j,\hat{Y}^{\rm qry}_j)\}_{j=1..K}$, where $K$ are hyper-parameters which control the size of meta-dataset.

We apply a two-level optimization on our prefix-tuning model for each project 
as show in Algorithm \ref{alg:cap}
. For inner optimization, we perform a NLL training object on $M(\phi)$ with support set $c^{\rm meta-sup}_i$:

\begin{equation}
\begin{split}
    \mathcal{L}_{\rm inner}(\phi, c^{\rm meta-sup}_i) &= \sum_{(X, \hat{Y}) \in c^{\rm meta-sup}} -\log p(\hat{Y}; X, \phi) \\
    & = \sum_{(X, \hat{Y}) \in c^{\rm meta-sup}_i} - \frac{1}{|\hat{Y}|}\sum^{|\hat{Y}|}_{t=1} Pr(y_t=\hat{y}_t|X, \phi)
\end{split}
\end{equation}

After the optimization, instead of directly updating model parameters $\phi$, we keep it still and use a copied parameter $\phi'$ to store the new value, which is adapted to project $\mathcal{P}_i$ as:

\begin{equation}
    \phi' = \phi - \lambda_{\rm inner} \nabla_\phi \mathcal{L}_{\rm inner}(\phi, c^{\rm meta-sup}_i).
\end{equation}

where $\lambda_{\rm inner}$ indicates learning rate of inner optimization process.

In the outer optimization, we perform the NLL training object on $M(\phi')$ with query set $c^{\rm meta-qry}_i$:

\begin{equation}
\begin{split}
    \mathcal{L}_{\rm outer}(\phi', c^{\rm meta-qry}_i) &= \sum_{(X, \hat{Y}) \in c^{\rm meta-qry}} -\log p(\hat{Y}; X, \phi') \\
    & = \sum_{(X, \hat{Y}) \in c^{\rm meta-qry}_i} -\frac{1}{|\hat{Y}|}\sum^{|\hat{Y}|}_{t=1} Pr(y_t=\hat{y}_t|X, \phi')
\end{split}
\end{equation}

But in the outer optimization, we consider $\phi'$ as value calculated from $\phi$, and optimize $\phi$ instead as following:

\begin{equation}
    \phi = \phi - \lambda_{\rm outer} \nabla_\phi \mathcal{L}_{\rm outer}(\phi', c^{\rm meta-qry}_i).
\end{equation}

where $\lambda_{\rm outer}$ indicates learning rate of outer optimization process.

After training with a meta project $\mathcal{P}_i$, the model not only adapt to the source project that construct $\mathcal{P}_i$, but also learns a better parameter $\phi$ which can adapt faster, making the model more generalized from different projects.

\begin{algorithm}
\caption{Prefix Meta-Learning}\label{alg:cap}
\begin{algorithmic}
\Require $p(\mathcal{T})$: Distribution of meta-tasks within source projects
\Require $\lambda_{\rm inner},\lambda_{\rm outer}$: learning rate hyper-parameters
\State 1. randomly initialize $\phi$
\State 2. \textbf{while} not converge \textbf{do}
\State 3. \, Sample batch of project $\mathcal{P}_i \sim p(\mathcal{T})$ 
\State 4. \, \textbf{for all} $\mathcal{P}_i$ \textbf{do}
\State 5. \, \, \textbf{for in range(inner\_steps) do}
\State 6. \, \, \, Evaluate $\mathcal{L}_{\rm inner}(\phi, c^{\rm meta-sup}_i)$ with respect to support set of this project
\State 7. \, \, \, Calculate adapted parameters with gradient descent: $\phi' = \phi - \lambda_{\rm inner} \nabla_\phi \mathcal{L}_{\rm inner}(\phi, c^{\rm meta-sup}_i)$
\State 8. \, \, \textbf{end for}
\State 9. \, \, Evaluate $\mathcal{L}_{\rm outer}(\phi', c^{\rm meta-qry}_i)$ with respect to query set of this project 
\State 10. \, Update $\phi = \phi - \lambda_{\rm outer} \nabla_\phi \mathcal{L}_{\rm outer}(\phi', c^{\rm meta-qry}_i)$ 
\State 11. \, \textbf{end for}
\State 12. \textbf{end while}
\end{algorithmic}
\end{algorithm}

\subsection{Meta Datasets Selection} \label{sec:approach_datasets}

Meta-transfer learning requires the creation of different meta tasks to increase generalization to new target projects. In applications such as classification ~\cite{DBLP:conf/cvpr/SunLCS19}, meta tasks can be easily defined using class labels from a single corpus. But this method does not work for code summarization. Since tasks can be defined as data from different projects, the problem becomes how to select suitable source projects, where each source project should have as many identities as possible similar to the target project. Following ~\cite{ChenS21a}, we consider the following metrics  may aid in source project selection: (1) code representation similarity, (2) cosine similarity, (3) length similarity, (4) ROUGE-2 recall, (5) ROUGE-2 accuracy (see \cite{ChenS21a} for more details). Through preliminary investigation, we select top 3 projects as our meta projects according to the average ranking of the following criteria to : (1) ROUGE-2 recall, (2) code representation similarity , and (3) code length.

%% file: experiment.tex
\section{Experiment Settings} \label{sec:experiment}

\subsection{Project-Specific Code Summarization Dataset}

We construct a new dataset to investigate the proposed low-resource project-specific code summarization task and evaluate the effectiveness of our method. Firstly, we select nine diversified active projects of Apache, Google, and Spring, as shown in Table \ref{tab:sourceprojects}. 

To make these projects a suitable dataset for code summarization, we extract all the methods in Java files after cloning their repositories. Then, following \cite{Leclair2019A}, we extract the first sentence by looking for the first period, or the first newline if no period was present as the corresponding summary. Next, to reduce noisy data, we clean the extracted code summary pairs of the following criteria: (1) summaries generated automatically by tools (e.g., IDE), (2) summaries whose length is less than three, (3) summaries are not in English. Finally, we curate a project-specific code summarization dataset consisting of more than 3.7k code-summary pairs. The numbers of code-summary pairs for each project are shown in Table \ref{tab:sourceprojects}.

\subsection{Principle behind the Dataset Construction}
\label{sec:princple}

To make the dataset as suitable as possible for low-resource code summarization evaluation, the project selection process above follows two main rationales.

(1) Projects should be diversified to guarantee the generalization of domain adaptation capabilities. We expect PCS models to face more complex and diverse transfer learning scenarios, instead of a group of similar projects implying similar domain knowledge. So we choose projects covering versatile domains from three different organizations.

(2) Selected projects should be well maintained to have enough high-quality golden code summaries to verify the model performance accurately. One might question why not use real poorly-documented projects for evaluation. Assuming we choose a project with only 10 code summaries for our 10-sample setting (only ten code-summary pairs for training), there will be no extra golden summaries for testing. One possible approach is to test with manually-annotated summaries, but this is labor-extensive and risky in introducing bias (e.g., unexpected inconsistency between manually-labeled summaries and existing ones). Constructing low-resource scenarios based on well-maintained projects may yield a relatively more convincing experimental setting even though it is not perfect. Indeed, it is also a common strategy to simulate data scarcity scenarios by using only a tiny part of off-the-shelf data for training (e.g., \cite{zhang2020PEGASUS} and \cite{ChenS21a}). 

Next we will introduce how to simulate low resource scenarios based on our dataset.

\begin{table}
\caption{Number of code-summary pairs in each project.}
\label{tab:sourceprojects}
\centering
\begin{tabular}{ccc}
\hline
Organization & Project  & \# Code-Summary Pairs \\
\hline
\multirow{3}*{Spring} & Spring-Boot & 2252 \\
 & Spring-Framework & 11020 \\
 & Spring-Security & 2092 \\
\multirow{3}*{Google} & Guava & 2122 \\
 & ExoPlayer & 2768 \\
 & Dagger & 928 \\
\multirow{3}*{Apache} & Kafka & 3453 \\
 & Dubbo & 1085 \\
 & Flink & 11551 \\
\hline
\end{tabular}
\end{table}

\subsection{Low-Resource Scenario Settings}

We simulate the low-resource scenario based on a 5-fold cross-validation process. Specifically, the code-summary pairs in each project are randomly divided into five equally-sized folds. At each cross-validation step, we use one fold as the testing set, and the other four folds as the candidate training set. Then, to construct the low-resource scenario, following \cite{ChenS21a}, we randomly select only 10 or 100 samples from the candidate training set for fine-tuning. So we have two low-resource scenarios, named 10-sample and 100-sample, respectively. 

\subsection{Implementation Detail}

\subsubsection{Hyperparameters}

Following Wasi et al.\cite{wasi2020transformer}, the number of layers and heads of transformer and the hidden size is set to 6, 8 and 512 respectively.
The vocabulary size of the code text, and document are 31945 and 28354. 
The maximum lengths for code sequence and document sequence are 150 and 30, covering at least 80\% of the initial set in project-specific code summarization dataset. Sequences that exceed the maximum will be truncated and the shorter sequence are padded with zeros. 
We remove the code-summary pairs existing in the selected nine projects from the pre-training dataset to avoid information leakage.
And To avoid the over-fitting problem when training large models in the low-resource scenario, we keep transformer parameters frozen except the last linear layer.
We train the our models for a maximum of 30 epochs and perform early stop if the validation performance does not improve for 5 consecutive iterations.
Adam is used for parameter optimization. The learning rate is set to 0.00005. Our models are implemented in PyTorch and trained on Tesla T4.

\subsubsection{Evaluation details}

For each target project, we use the rest of the projects as candidates to build meta-datasets. The combination of source projects in the meta-dataset is decided according to the proposed criteria described in section \ref{sec:approach_datasets}. Following \cite{ChenS21a}, we fine-tune the meta learned model with 10 or 100 labeled samples on the target project.
And to prevent randomness during low-resource training, the project-specific fine-tuning process was repeated 10 times and the average result was used as the result of one cross-validation step.

\subsection{Evaluation Metrics}

Following previous works \cite{ChenS21a}, we evaluate the performance of GCS and PCS models based on BLEU \cite{Papineni2002BLEU} and ROUGE-L\cite{lin2004rouge}, all of which are widely used in evaluating performance of text generation tasks. BLEU score is a popular accuracy-based measure for machine translation, as well as in code summary generation tasks. It calculates the similarity between the generated sequence and reference sequence by counting the n-grams that appear in both the candidate sequences and the reference sequence. ROUGE-L takes into account sentence level structural similarity naturally and identifies longest co-occurring in sequence n-grams automatically.

\subsection{Baselines}

To evaluate the effectiveness of our model, we compare our approach with existing work on source code summarization and text summarization. They can be divided into two groups: GCS models and Transfer-learning-based (Transfer learning-based) models.

\subsubsection{GCS models}

\begin{itemize}

\item \textbf{CodeNN:} CodeNN\cite{Iyer2016Summarizing} is the first end-to-end code summarization approach. They use LSTM to generate summaries given code snippets. 

\item \textbf{AST-AttendGRU:} AST-AttendGRU\cite{Leclair2019A} is an approach which uses both original code text and SBT representation as input. The approach builds a GRU-based-encoder-decoder model with two encoders and one decoder. 

\item \textbf{Projcon:} Projcon \cite{aakash2021projcon} is a GCS model which integrate contextual information into code summarization. Projcon creates a vectorized representation of selected code files in a project, and use that representation to augment the encoder of SOTA neural code summarization techniques. 

\item \textbf{Transformer:} Transformer \cite{vaswani2017attention} is the most popular NMT architecture, which consists of stacked multi-head attention and parameterized linear transformation layers for both the encoder and decoder.

\item \textbf{Rencos:} Rencos \cite{gregg2020rencos} is a retrieval-based neural model that augments an attentional encoder-decoder model with the retrieved two most similar code snippets for better source code summarization.

\item \textbf{EditSum:} EditSum \cite{li2021EditSum} is also a retrieval-based neural model which first retrieves a similar code snippet as a prototype summary and then edits the prototype automatically with the semantic information of input code.

\end{itemize}

\subsubsection{Transfer-learning-based models}

\begin{itemize}

\item \textbf{Adaptor:}  Adaptor \cite{neil2019adapter} is a transfer learning method, which insert adapter layers after each feed-forward layer of the transformer model, which reduces the number of parameters that needs to be trained and prevent over-fitting.

\item \textbf{MAM-Adaptor:} MAM-Adaptor \cite{he2021towards} is a unified-model that improves on adapter methods, which learns a modification to hidden representation of model.
    
\item \textbf{MTL-ABS:} MTL-ABS \cite{ChenS21a}  is a meta-transfer learning based text abstractive summarization model which leverage multiple corpora to improve the quality of generated summary.

\end{itemize}

%% file: results.tex
\section{Experimental Results}

This section evaluates the effectiveness of our MPCos and investigates the proposed low-resource project-specific learning task. Specifically, we aim to answer the following research five questions:

\begin{itemize}

\item \textbf{RQ1:} How does MPCos perform compared to state-of-the-art GCS baselines?
\item \textbf{RQ2:} How does MPCos perform compared to transfer-learning-based baselines?
\item \textbf{RQ3:} How does meta-learning affect the overall performance of MPCos?
\item \textbf{RQ4:} How does the prefix-tuning affect the overall performance of MPCos?
\item \textbf{RQ5:} How is project-specific knowledge learned in MPCos?

\end{itemize}  

\subsection{RQ1: MPCos v.s. GCS models}\label{ses:gcs_results}

\begin{table}[t]
  \caption{
Experimental results of our model compared with GCS models. 
``$^*$'' means the result is adapted from the original papers and ``-'' means the corresponding result is unavailable. 
Note that we are not listing the performances of GCS and PCS scenarios to compare them. Indeed, it is not a fair comparison. Instead, we mainly emphasize the promising research opportunity of introducing project-specific knowledge based on GCS models. 
  }
  \label{tab:performance_gcs_in_project}
  \centering
  \begin{tabular}{llrrr}
  \hline
   Scenario & Approach & BLEU & ROUGE\_L\\ 
   \hline
   \multirow{6}*{GCS}&
   CodeNN         &  9.95$^*$ & - \\
  &AST-AttendGRU  & 19.60$^*$ & - \\
  &Projcon        & 17.45$^*$ & 50.37$^*$ \\
  &Transformer    & 44.58$^*$ & 54.76$^*$ \\
  &Rencos         & 27.35$^*$ & 42.00$^*$ \\
  &EditSum        & 28.06$^*$ & 53.17$^*$ \\
    \hline
   \multirow{8}*{PCS}&
   CodeNN         & 3.19 & 22.32 \\
  &AST-AttendGRU  & 4.33 & 25.25 \\
  &Projcon        & 4.13 & 25.75 \\
  &Transformer    & 4.77 & 26.28 \\
  &Rencos         & 4.41 & 25.73 \\
  &EditSum        & 4.83 & 26.45 \\
  &MPCos(10)      & 7.03 & 30.63 \\
  &MPCos(100)     & \textbf{9.10} & \textbf{33.21} \\
    \hline
\end{tabular}
\end{table}

\begin{table*}[tb]
  \caption{
  Experimental results of our model compared with Transfer-learning-based models. 
  The improving ratios column are calculated for MPCos over MTL-ABS, which is the SOTA model on low resource text summarization. 
  }
  \label{tab:overall_performance}
  \centering
  \begin{tabular}{ccccccc||cc}
  \hline
    \multirow{2}*{Dataset}& Labeled & Adaptor & MAM-Adaptor & MTL-ABS & MPCos & Improving ratio & MPCos w/o prefix & MPCos w/o meta \\
    & Samples & $B/R_L$ & $B/R_L$ & $B/R_L$ & $B/R_L$ & $B/R_L$ & $B/R_L$ & $B/R_L$ \\
\hline

 Spring& 10 & 2.88/19.63 & 3.58/20.27 & 5.69/29.60 & 8.21/\textbf{34.50} & 44.14\%/16.57\% & \textbf{8.31}/33.67 & 6.80/31.75 \\
 Boot& 100 & 2.88/19.63 & 3.81/20.53 & 8.15/33.54 & \textbf{11.96}/\textbf{37.79} & 46.73\%/12.67\% & 9.91/35.36 & 8.87/34.55 \\
 Spring& 10 & 2.47/19.74 & 3.40/20.60 & 5.49/29.40 & 8.08/\textbf{32.45} & 47.19\%/10.36\% & \textbf{8.29}/\textbf{32.41} & 7.09/30.86 \\
 Framework& 100 & 3.99/23.66 & 3.52/20.97 & 8.54/34.28 & \textbf{9.49}/\textbf{35.78} & 11.17\%/4.38\% & 9.36/34.27 & 9.39/34.82 \\
 Spring& 10 & 2.13/18.88 & 2.86/18.82 & 3.23/25.15 & 4.77/\textbf{26.98} & 47.58\%/7.25\% & \textbf{4.90}/26.25 & 4.14/26.12 \\
 Security& 100 & 2.11/18.75 & 3.17/20.18 & 4.42/27.83 &\textbf{ 5.96}/\textbf{28.75} & 34.77\%/3.32\% & 5.60/28.10 & 5.08/27.48 \\
\multirow{2}*{Guava}
 & 10 & 2.92/19.17 & 4.10/20.13 & 5.63/27.96 & 7.71/31.30 & 36.94\%/11.97\% & \textbf{9.72}/\textbf{31.69} & 7.33/30.58 \\
 & 100 & 3.07/19.27 & 4.06/19.85 & 7.55/31.54 & \textbf{13.12}/\textbf{34.69} & 73.68\%/9.98\% & 10.97/33.53 & 8.39/32.67 \\
\multirow{2}*{ExoPlayer}
 & 10 & 1.36/19.08 & 2.37/19.95 & 6.67/31.84 & \textbf{8.49}/\textbf{34.42} & 27.36\%/8.08\% & 8.07/33.69 & 7.76/32.83 \\
 & 100 & 1.36/19.18 & 2.57/20.27 & 7.93/34.45 & \textbf{9.59}/\textbf{36.69} & 20.85\%/6.50\% & 9.17/35.40 & 9.47/35.78 \\
\multirow{2}*{Dagger}
 & 10 & 2.74/18.50 & 3.57/20.01 & 5.88/32.15 & \textbf{8.82}/35.33 & 50.05\%/9.88\% & 7.89/34.68 & 8.15/\textbf{36.52} \\
 & 100 & 2.98/19.01 & 3.95/19.89 & 7.62/35.80 & 11.21/\textbf{40.41} & 47.16\%/12.87\% & 10.20/37.09 & \textbf{11.35}/37.34 \\
\multirow{2}*{Kafka}
 & 10 & 1.06/15.69 & 2.16/16.82 & 3.42/25.16 & \textbf{5.14}/\textbf{27.05} & 50.46\%/7.50\% & 4.81/26.82 & 4.60/26.84 \\
 & 100 & 1.20/16.14 & 2.18/16.89 & 4.39/26.45 & \textbf{6.10}/\textbf{28.46} & 38.71\%/7.57\% & 5.18/27.37 & 5.41/27.60 \\
\multirow{2}*{Dubbo}
 & 10 & 2.24/13.14 & 1.00/14.02 & 4.22/24.18 & \textbf{6.00}/\textbf{24.90} & 42.22\%/2.96\% & 4.89/23.88 & 5.62/24.69 \\
 & 100 & 3.14/13.31 & 1.00/14.12 & 4.52/25.46 &\textbf{ 7.63}/\textbf{26.56} & 68.91\%/4.31\% & 5.42/24.77 & 5.89/25.85 \\
\multirow{2}*{Flink}
 & 10 & 1.86/17.47 & 2.82/18.23 & 4.62/27.61 & \textbf{6.04}/28.72 & 30.52\%/4.04\% & 5.68/28.53 & 5.92/\textbf{29.13} \\
 & 100 & 2.14/18.55 & 2.76/18.23 & 4.89/28.03 & \textbf{6.88}/\textbf{29.75} & 40.62\%/6.15\% & 6.34/28.97 & 6.21/29.22 \\
\multirow{2}*{Average}
 & 10 & 2.18/17.92 & 2.87/18.76 & 4.98/28.12 & \textbf{7.03}/\textbf{30.63} & 41.78\%/8.68\% & 6.95/30.18 & 6.38/29.92 \\
 & 100 & 2.54/18.61 & 3.00/18.99 & 6.45/30.82 & \textbf{9.10}/\textbf{33.21} & 42.47\%/7.48\% & 8.02/31.65 & 7.78/31.70 \\
 
    \hline
\end{tabular}
\end{table*}

Our first research question is: How does MPCos perform compared to state-of-the-art GCS baselines? 
To answer this question, we compare our MPCos with the other six GCS baselines on both GCS and PCS scenarios.
First, in GCS Scenario, we use the best results reported in their original paper. 
Second, to obtain the results of GCS models in the PCS scenario, we first trained GCS models on the GCS datasets provided by \cite{Leclair2019recommendations}, and then evaluated them on our PCS datasets. 
Third, we train our MPCos model on a PCS scenario with 10 and 100 examples respectively, which are named MPCos(10) and MPCos(100). 
We evaluated these models on nine projects of PCS datasets, and the average results are reported due to page limitations. 
Looking into TABLE \ref{tab:performance_gcs_in_project}, we have two interesting observations.

First, The performance of all GCS models on our PCS dataset demonstrates a drastic degradation compared with their scores initially reported. For example, the BLEU of EditSum, the best-performance GCS model, on its own dataset is 28.06, while the score on the PCS dataset is only 4.83.

Second, By learning limited code-summary pairs (10 and 100), MPCos achieves substantial improvement over GCS models (e.g., nearly 100\% relative enhancement of BLEU compared with EditSum).

It is noteworthy that it is unfair to compare the performance between PCS and GCS scenarios directly, since PCS and GCS target different scenarios, and used different hyperparameters and datasets. However, we do not want to emphasize MPCos's performance superiority over GCS models, instead, we aim to demonstrate the promising opportunity of introducing project-specific knowledge based on GCS models. 

\subsection{RQ2: MPCos vs. Transfer-learning-based models} \label{sec:TL_results}

Our second research question is: How does MPCos perform compared to transfer-learning-based baselines? To answer this question, we compare another three transfer learning-based models (Adaptor, MAM-Adaptor, and MTL-ABS) on the nine target projects based on average BLEU and Rouge-$L$. The left part of Table \ref{tab:overall_performance} shows the performance of the baseline models and MPCos. By investigating the results, we have the following observations.

First, to our surprise, Adaptor and MAM-Adaptor, two robust parameter efficient fine-tuning models for text summarization, underperform in the PCS scenario. MTL-ABS, a recent low-resource abstractive text summarization model, demonstrates superiority over Adaptor and MAM-Adaptor, due to its low-resource learning capability. However, we can still observe a notable performance gap between MTL-ABS and our MPCos. One possible reason is that code summarization models are more sensitive to the input than text summarization, and our tuning method can capture more subtle differences in source code by better coordinating low-resource learning with knowledge of pre-trained models. A more detailed discussion will be presented in section \ref{sec:analysis_prefix}.

Second, MPCos of the 10-sample setting (MPCos(10)) have achieved remarkable improvements compared to Transformer, and further impressive performance boost can be observed when learning with 100 samples. For example, there are 30\% relative enhancements of BLEU score from MPCos(10) to MPCos(100)) on average. These results verify the overall effectiveness of our meta transfer learning process. We will investigate in-depth how project-specific knowledge is learn as the number of samples increases in section \ref{sec:analysis}.

\subsection{RQ3: Effect of meta learning} \label{sec:analysis_meta}

Our third research question is: How does meta-learning affect the performance of MPCos? To answer RQ3, we compare the performance of MPCos with \textbf{MPCos w/o meta}, which keeps only the prefix module by excluding the meta-training process from MPCos. The detailed evaluation results of the nine target projects are summarized in the right part of Table \ref{tab:overall_performance}, and we find both BLEU and Rouge-$L$ of \textbf{MPCos w/o meta} decrease in all projects. Overall, the BLEU and Rouge-$L$ decreased from 7.03/30.63 to 6.38/29.92 and 9.10/33.21 to 7.78/31.70 on the 10- and 100-sample scenarios, respectively. The results indicate that (1) source projects are valuable information for learning a high-quality code summarization model, especially from a project with few historical code summaries, and (2) meta-learning can effectively leverage this information.

Besides, we find meta-learning can facilitate gradient updates of fine-tuning. For example, the number of epochs to converge of \textbf{MPCos w/o meta} and MPCos is 14.48 and 18.70.

\subsection{RQ4: Effect of prefix-tuning} \label{sec:analysis_prefix}

Our fourth research question is: How does prefix-tuning affect the performance of MPCos? To answer RQ4,  We demonstrate the effects of prefix-tuning in Table \ref{tab:overall_performance} by removing the prefix component from MPCos (\textbf{MPCos w/o meta}). As shown in Table \ref{tab:overall_performance}, MPCos achieves better performance in most projects compared to \textbf{MPCos w/o prefix}. Specifically, comparing \textbf{MPCos w/o prefix} with MPCos, the BLEU and Rouge-$L$ decreased from 9.10/33.21 to 8.02/31.65 in the 100-sample scenario, which demonstrates the effectiveness of prefix tuning. However, \textbf{MPCos w/o prefix} achieves a similar performance with MPCos on the scenario of 10 labeled samples. The reason may be that with the help of meta-learning, \textbf{MPCos w/o prefix} can learn textual style information well, but need more samples (e.g., 100 samples) to understand deeper semantics.

Then we compare MTL-ABS with MPCos to further demonstrate the effectiveness of prefix-tuning. Both MTL-ABS and MPCos models freeze most pre-trained parameters and only tune a smaller set of parameters. The main difference between them is that MTL-ABS inserts additional adapters between the layers of pre-trained models, while our model inserts a sequence of continuous project-specific vectors to the input, which modifies the pre-trained model slightly than MTL-ABS. However, MPCos still remarkably outperforms MTL-ABS. 
The reason may be that code summarization is more sensitive to the input than text summarization. In other words, the code summary can diversify with minor modifications to the code. In this case, modifying the architecture of the pre-trained model may not well retain its learned knowledge. 

We verify this assumption with a simple experiment. Specifically, we initialize the parameters of vanilla Transformer layers with the pre-trained model, randomly initialize the other parts of MTL-ABS and MPCos, and evaluate them directly. The experimental results show that the performances of MTL-ABS and MPCos decrease compare to their base pre-trained model. However, the degradation of MPCos is much slighter than that of MTL-ABS (e.g., 28\% v.s. 61\% of relative decrement in terms of BLEU on the Guava project), indicating the correctness of our assumption. Thus, we can conclude that prefix-tuning is a more applicable fine-tune approach in our scenario.

\subsection{RQ5: How project-specific knowledge be learned}\label{sec:analysis}

\begin{figure}[t]
\includegraphics[width=1.0\linewidth]{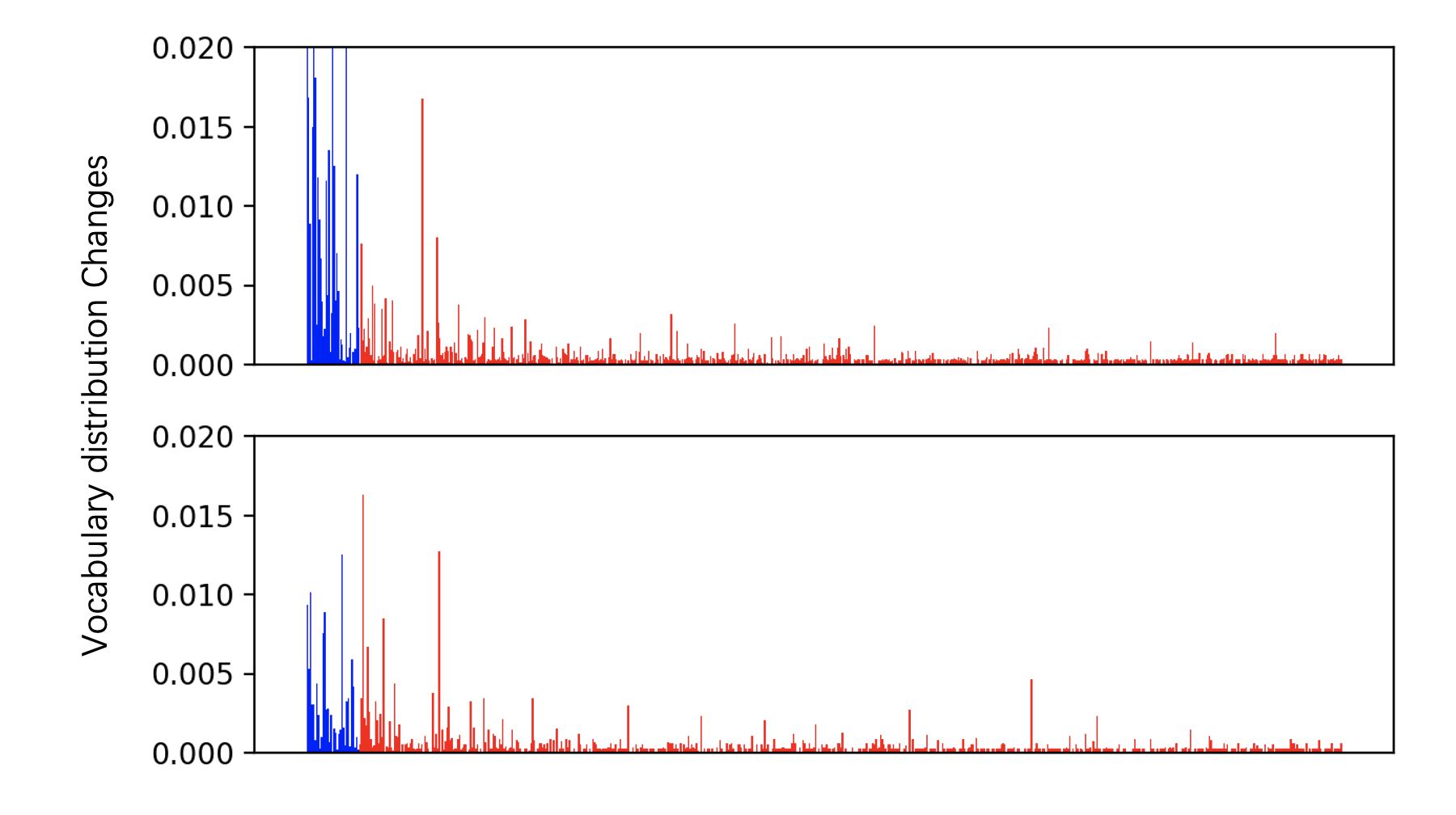}
\caption{
Word distribution changes of MPCos(10) over Transformer in the preliminary stage (above) and MPCos(100) over MPCos(10) in the refining stage (below). The X-axis represents summary words ordered by global word frequency from high to low. Y-axis represents the frequency changes in generated
summaries of each word. 
Blue bars represent high-frequency words, which are the top 5\% most frequent words according to Funcom dataset\cite{Leclair2019recommendations}, and red bars represent the other words.
}
\label{fig:distribution_changes}
\end{figure}

\begin{table}[t]
\caption{Generated summaries by Transfomer, MPCos(10), and MPCos(100) given the same input code. The key differences are marked in red.
}
\label{tab:style}
\centering
\begin{tabular}{ll}
\hline
\textbf{Examples 1} \\
\hline
\textbf{Reference}: returns an instance that escapes special \\
characters in a string so it can safely be included in xml \\
document as an attribute value \\ 
\textbf{Transformer}: an xml attribute which wraps the value \\ 
of the attribute \\
\textbf{MP(10)}: \textcolor{red}{returns} an xml attribute for this element \\
\textbf{MP(100)}: returns \textcolor{red}{an instance that escapes special} \\
\textcolor{red}{characters} in an xml document \\
\hline
\textbf{Examples 2} \\
\hline
\textbf{Reference}: returns the method instance for so that tests \\of java \\ 
\textbf{Transformer}: a very custom way to add a null method \\
\textbf{MP(10)}: \textcolor{red}{returns the} add \textcolor{red}{method} to use for adding\\ null values \\
\textbf{MP(100)}: returns the method \textcolor{red}{instance so tests can suppress it} \\
\hline
\end{tabular}
\end{table}

Our fifth research question is: How is project-specific knowledge learned in MPCos? To answer RQ5, we look into the internal process of learning project-specific knowledge, revealing how summaries evolve with the increment of learning samples. 

Specifically, we manually split the learning process into two stages, including preliminary and refining ones. By saying preliminary stage, we mean the learning process of MPCos with ten examples upon the pre-trained Transformer (from Transformer to MPCos(10)). And the refining stage refers to the further learning process of MPCos with 100 examples (from MPCos(10) to MpCos(100)).

We use word distribution of the generated code summaries to investigate how they evolve in these two learning stages. Each word will be assigned a real number, representing the frequency of its occurrence. We also have a global word distribution (named vocabulary distribution as a distinction) over the large-scale Funcom dataset provided by LeClair et al \cite{Leclair2019recommendations}.  According to vocabulary distribution, we can distinguish high-frequency and low-frequency words , which roughly correspond to project-specific writing-style words (or patternized words) and domain topic words.

We then draw two histograms for the two learning stages to reveal how summaries evolve with the increment of learning samples. 
The preliminary stage is expressed by Figure \ref{fig:distribution_changes} (above). 
The X-axis represents summary words ordered by vocabulary distribution, and their global frequencies decrease from left to right. 
Y-axis represents the frequency changes in generated summaries of each word. 
Blue bars represent high-frequency words, and red bars represent the other words.

As shown in Figure \ref{fig:distribution_changes} (above), most of the frequency changes occur at the left positions, indicating the general words (e.g., prepositions, pronouns, or keywords in Java) may change enormously from summaries generated by Transformer to ones generated by MPCos(10). Thus, we can safely conclude that our model mainly learns the shallow knowledge of the textual style in a project by training with the ten examples.

We show some concrete cases in Table \ref{tab:style}. In the first example, the summaries generated by Transformer and MPCos(10) are both consistent with the human-written ones in terms of semantics. However, MPCos(10)'s output matches the golden summary better in textual style. The textual style is mainly expressed via the high-frequency words like ``return'' and ``an'', and MPCos(10) successfully learns to generate these words.

Regarding the refining stage, we can find in Figure \ref{fig:distribution_changes} (Below) that word frequency changes are much more balanced across the X-axis. This trend suggests MPCos(100) captures more profound domain knowledge of source code by learning more examples. Recapping the first example in Table \ref{tab:style}, MPCos(100) generates a better summary, which is not only consistent with the golden summary in textual style, but also fixes the factual error generated by Transformer and MPCos(10), e.g., by changing ``an xml attribute'' to more accurate content of ``an instance that escapes special characters''. 

Generally,  this experiment reveals that as the number of training examples increases, MPCos learns project-specific knowledge from shallow textual features to deep semantic information, progressively.

\input{human-evaluation}

%% file: human-evaluation.tex
\subsection{Human Evaluation}

\begin{table}[t]
  \caption{
  Experimental results of human evaluation.
  We evaluated these models on nine projects of PCS datasets, and the average results are reported considering page limitation.
  }
  \label{tab:human_evaluation}
  \centering
  \begin{tabular}{lcccc}
  \hline
   Approach & Naturalness & Informativeness & Relevance\\ 
  \hline
Transformer & 3.65 & 3.27 & 2.75 \\
EditSum     & 3.78 & 3.43 & 3.10 \\
MTL-ABS(10) & 3.87 & 3.48 & 3.07 \\
MTL-ABS(100)& 3.84 & 3.72 & 3.48 \\
MPCos(10)   & 3.90 & 3.62 & 3.44 \\
MPCos(100)  & 3.96 & 3.85 & 3.81 \\
    \hline
  
\end{tabular}
\end{table}

Since the automatic metrics we used mainly focus on text similarities, especially the lexical overlap between generated summary and the human-written summary, they can not reasonably reflect the similarity at the semantic level. 
Therefore, we perform a human evaluation to measure the semantic similarities of the output of MPCos and baselines.
Following previous works \cite{Iyer2016Summarizing, li2021EditSum}, the naturalness (grammaticality and fluency of the generated summaries) and informativeness (the amount of content carried over from the input code to the generated summaries, ignoring fluency of the text) scores (ranging from 0 to 4, from bad to good) would be asked to evaluate the generated summaries.
Furthermore, another human evaluation metric, relevance, was designed to measure whether the generated summary is relevant to the input code.

We invited ten volunteers with relevant experience in software development and good English proficiency to score the summaries generated by each model in the test set. They are doctoral or master students majoring in software engineering, and they are not co-authors of this work. For each baseline that needs to be evaluated, we pick 50 samples in each project's test set, mix the results generated by different baselines, and hide their provenance. They were then assigned to different volunteers, and we ensured that at least three volunteers evaluated each sample. The final evaluation score was the average of all the volunteers' scores.

The evaluation results are shown in the Table \ref{tab:human_evaluation}, from which we have the following observations: 

First, MPCos outperforms all baselines in terms of naturalness, informativeness, and relevance, indicating the effectiveness of our model.

Second, Compared MPCos with GCS models, the improvement on naturalness is much slighter than that on informativeness and relevance, showing that (1) the effectiveness of GCS models in capturing cross-project general knowledge, and (2) project-specific knowledge is also a valuable source to generate high-quality code summarization.

Finally, MPCos(100) achieves similar performance with MPCos(10) on naturalness, but significantly improves informativeness and relevance. This observation verifies our previous finding that the PCS model progressively learns project-specific knowledge from shallow textual features to deep semantic information.

%% file: discussion.tex
\section{Discussion}

\subsection{Time cost of Meta-Prefix tuning}

To evaluate the practical feasibility of our method, we analyze the training time cost and inference time cost. 
Note that MPCos have an extra meta-tuning and a project-specific tuning process compared with GCS models. 
When training MPCos on nine projects, the average cost time of meta-tuning and project-specific tuning is 148 seconds (14.8 epochs times 10 seconds/epochs) and 99 seconds (16.5 epochs times 6 seconds/epochs), respectively. 
The extra training time is short enough compared to the code revision period in a real-world code summarization scenario. 
Regarding the inference, since the parameter size between MPCos and GCS is similar, their inference time cost is generally the same, which is 102.4 samples/second on average. 
Most importantly, once the GCS model is pre-trained, we only need to perform lightweight transfer learning upon it, eliminating the time-expensive training on large-scale multi-project GCS datasets later. 
Therefore, we can conclude that our meta-transfer learning method can yield a feasible solution for project-specific code summarization in terms of efficiency.

\subsection{Directions of Future Research}\label{sec:future}

Project-specific code summarization is still far from being fully explored as a practical and novel task in the software engineering community. We summarize future directions as follows.

\begin{itemize}

\item \textbf{Context Extraction}. According to our experimental results, the project-specific information greatly affects code summary generation. In this paper, we use a meta-prefix tuning method to extract the implicit knowledge. We believe the explicit knowledge from a specific project may be another valuable source for code summarization.

\item \textbf{Zero-shot PCS scenario}.
In a low-resource PCS scenario, we have limited code-summary pairs as extra training data to adapt project-specific characteristics. In a more extreme case, there can be no historical code summaries. This case is referred to as the zero-shot PCS scenario. A straightforward strategy is to convert the zero-shot PCS task into a typical low-resource PCS task via manually writing a certain amount of code summaries. Thus, which methods should be manually summarized first and how the method selection method affects the performance of code summarization are interesting research questions to investigate.


\end{itemize}

\subsection{Threats to Validity}

There are three main threats to the validity of our evaluation.

\begin{itemize}

   \item The first threat is that our experiment results may apply only to the nine projects collected on our PCS datasets. To reduce this threat,  as described in \ref{sec:princple}, we have employed a principled project selection method. However, in any case, the nine selected projects can not represent all practical scenarios of working on multiple projects. Project selection is a nontrivial problem and also deserves future research.

    \item The second threat is that we use the first sentence of code documentation as our ground truth. This is also a common practice in prior research efforts on code summarization \cite{Leclair2019A}. This approach will inevitably introduce noise into the data, e.g., mismatches between methods and the one-sentence summarization. 

    \item The third threat is that we re-implement the EditSum and report its result based on our re-implement version. Our implementation details may differ from the original work, though we have tried our best to read the paper carefully.

\end{itemize}

%% file: related_work.tex
\section{Related Work}

\subsection{General Code Summarization}

Automatic code summarization now is an important and rapidly-growing research topic in the community of software engineering and natural language processing. For traditional techniques,  we direct refers to a comprehensive survey by Nazar et al. ~\cite{DBLP:journals/jcst/NazarHJ16}. Since the first neural model of code summarization was proposed by Iyer et al. \cite{Iyer2016Summarizing}, we have witnessed the introduction of the latest neural technologies for text generation tasks (e.g., machine translation and text summarization) into this research field recently. For example, Wei et al.~\cite{DualCos} and Ye et al. ~\cite{ye2020dual} introduced the code generation task to improve code summarization task via dual learning. 
Both Yang et al.~\cite{DBLP:conf/iwpc/YangKYGWMZ21} and Lin et al. ~\cite{DBLP:conf/iwpc/LinOZCLW21} present the AST based method to leverage structure information to improve code summarization.

Besides content in code snippets, project contextual information is also an important feature in code summarization, Bansal et al. ~\cite{DBLP:conf/iwpc/BansalHM21} proposed a project-level encoder to improve models of code summarization. which creates vectorized representation of selected code files in a software project, and use that representation to augment the encoder of state-of-the-art neural code summarization techniques. But their method only uses contextual information as features, although it can supplement project information to a certain extent, the information utilization rate is low, and it is difficult for the model to infer the style and domain of the current code summary from the context.

\subsection{Parameter Efficient Learning}

To improve the speed and stability of training, and prevent model from over-fitting to bias of certain dataset, parameter efficient learning method are proposed to reduce the number of training parameters. 
Houlsby et al. ~\cite{DBLP:conf/icml/HoulsbyGJMLGAG19} proposed a Adapter module that insert after each Feed-Forward layer of transfomers and freeze other parameters during training. 
Stickland et al. ~\cite{DBLP:conf/icml/Stickland019} developed a Projected Attention Layers(PALs) for multi-task natrual language understand, which are small layers with a similar role to Adapters. 
He et al. \cite{he2021towards} further proprosed a unified-model that improves on adapter methods, which learns a modification to hidden representation of model.
Our method follows recent prefix tuning method ~\cite{DBLP:conf/acl/LiL20}, which insert prefix vectors into each layer of transformers. Comparing to \cite{DBLP:conf/icml/HoulsbyGJMLGAG19} and \cite{he2021towards}, our method can capture the project-specific knowledge more effective and efficient.




\subsection{Meta-Transfer Learning}

Transfer learning techniques are often used to tackle the limited resource problem.
Finn et al. ~\cite{DBLP:conf/icml/FinnAL17} proposed a Modal-Agnostic Meta-Learning(MAML) method, that teach model how to learn in very limited resource by given modal a good initialization parameter which can fast adapt to different task or domains.
Sun et al. ~\cite{DBLP:conf/cvpr/SunLCS19} further improve MAML by introducing adapter into model, which decrease the difficulty to train a model in MAML paradigm and proposed meta-transfer learning.
Chen~\cite{ChenS21a} proposed to solve low-resource text summarization problem with the help of meta-transfer learning. To better fit the text summarization tasks, they investigate the problem of choosing source corpora for meta-dataset and provide some general criteria to mitigate the negative effect from inappropriate choices of source corpora.
Our method follows work of Chen~\cite{ChenS21a}, but the performance of the adapter is poor in the field of code summarization task. We further improve it with the help of prefix-tuning.




%% file: conclusion.tex
\section{Conclusion}

We have introduced the task of low-resource project-specific code summarization and presented our method that addresses the critical challenges of this task. The novelty of the proposed task lies in that developers are more concerned with generating high-quality summaries for their working projects, especially those with limited historical code summaries, while current code summarization efforts fail to investigate their effectiveness on specific projects. To solve this more practical task, our method effectively captures project knowledge from limited training samples by systematically integrating a project-specific prefix-based fine-tuning mechanism into a meta-transfer learning framework.  This simple yet effective method can serve as a solid baseline for future low-resource project-specific code summarization studies. We also discuss open questions and research opportunities and will release our source code and dataset soon\footnote{https://github.com/pkuserc/MPCos\_ASE2022}.